\documentclass[%
 aip,
 jmp,%
 amsmath,amssymb,
 superscriptaddress,
 reprint,%
]{revtex4}

\usepackage{amsmath,amsfonts,amssymb,color,epsfig,graphics,graphicx,latexsym,revsymb,theorem,url,verbatim,subfigure}

\newtheorem{definition}{Definition}
\newtheorem{proposition}[definition]{Proposition}
\newtheorem{lemma}[definition]{Lemma}

\newtheorem{theorem}[definition]{Theorem}
\newtheorem{corollary}[definition]{Corollary}
\newtheorem{conjecture}[definition]{Conjecture}

\newtheorem{remark}[definition]{Remark}
\newtheorem{example}[definition]{Example}
\newtheorem{question}[definition]{Question}

\newenvironment{proof}{\noindent \textbf{{Proof.~} }}{\qed}
\def\Dbar{\leavevmode\lower.6ex\hbox to 0pt
{\hskip-.23ex\accent"16\hss}D}
\makeatletter
\def\url@leostyle{%
  \@ifundefined{selectfont}{\def\UrlFont{\sf}}{\def\UrlFont{\small\ttfamily}}}
\makeatother
\def\bcj{\begin{conjecture}}
\def\ecj{\end{conjecture}}
\def\bcr{\begin{corollary}}
\def\ecr{\end{corollary}}
\def\bd{\begin{definition}}
\def\ed{\end{definition}}
\def\bea{\begin{eqnarray}}
\def\eea{\end{eqnarray}}
\def\bem{\begin{enumerate}}
\def\eem{\end{enumerate}}
\def\bex{\begin{example}}
\def\eex{\end{example}}
\def\bim{\begin{itemize}}
\def\eim{\end{itemize}}
\def\bl{\begin{lemma}}
\def\el{\end{lemma}}
\def\bpf{\begin{proof}}
\def\epf{\end{proof}}
\def\bpp{\begin{proposition}}
\def\epp{\end{proposition}}
\def\bqu{\begin{question}}
\def\equ{\end{question}}
\def\br{\begin{remark}}
\def\er{\end{remark}}
\def\bt{\begin{theorem}}
\def\et{\end{theorem}}

\def\btb{\begin{tabular}}
\def\etb{\end{tabular}}

\newcommand{\nc}{\newcommand}

 \nc{\bA}{{\bf A}} \nc{\bB}{{\bf B}} \nc{\bC}{{\bf C}}
 \nc{\bD}{{\bf D}} \nc{\bE}{{\bf E}} \nc{\bF}{{\bf F}}
 \nc{\bG}{{\bf G}} \nc{\bH}{{\bf H}} \nc{\bI}{{\bf I}}
 \nc{\bJ}{{\bf J}} \nc{\bK}{{\bf K}} \nc{\bL}{{\bf L}}
 \nc{\bM}{{\bf M}} \nc{\bN}{{\bf N}} \nc{\bO}{{\bf O}}
 \nc{\bP}{{\bf P}} \nc{\bQ}{{\bf Q}} \nc{\bR}{{\bf R}}
 \nc{\bS}{{\bf S}} \nc{\bT}{{\bf T}} \nc{\bU}{{\bf U}}
 \nc{\bV}{{\bf V}} \nc{\bW}{{\bf W}} \nc{\bX}{{\bf X}}
 \nc{\bZ}{{\bf Z}}

\nc{\cA}{{\cal A}} \nc{\cB}{{\cal B}} \nc{\cC}{{\cal C}}
\nc{\cD}{{\cal D}} \nc{\cE}{{\cal E}} \nc{\cF}{{\cal F}}
\nc{\cG}{{\cal G}} \nc{\cH}{{\cal H}} \nc{\cI}{{\cal I}}
\nc{\cJ}{{\cal J}} \nc{\cK}{{\cal K}} \nc{\cL}{{\cal L}}
\nc{\cM}{{\cal M}} \nc{\cN}{{\cal N}} \nc{\cO}{{\cal O}}
\nc{\cP}{{\cal P}} \nc{\cQ}{{\cal Q}} \nc{\cR}{{\cal R}}
\nc{\cS}{{\cal S}} \nc{\cT}{{\cal T}} \nc{\cU}{{\cal U}}
\nc{\cV}{{\cal V}} \nc{\cW}{{\cal W}} \nc{\cX}{{\cal X}}
\nc{\cZ}{{\cal Z}}

\nc{\hA}{{\hat{A}}} \nc{\hB}{{\hat{B}}} \nc{\hC}{{\hat{C}}}
\nc{\hD}{{\hat{D}}} \nc{\hE}{{\hat{E}}} \nc{\hF}{{\hat{F}}}
\nc{\hG}{{\hat{G}}} \nc{\hH}{{\hat{H}}} \nc{\hI}{{\hat{I}}}
\nc{\hJ}{{\hat{J}}} \nc{\hK}{{\hat{K}}} \nc{\hL}{{\hat{L}}}
\nc{\hM}{{\hat{M}}} \nc{\hN}{{\hat{N}}} \nc{\hO}{{\hat{O}}}
\nc{\hP}{{\hat{P}}} \nc{\hR}{{\hat{R}}} \nc{\hS}{{\hat{S}}}
\nc{\hT}{{\hat{T}}} \nc{\hU}{{\hat{U}}} \nc{\hV}{{\hat{V}}}
\nc{\hW}{{\hat{W}}} \nc{\hX}{{\hat{X}}} \nc{\hZ}{{\hat{Z}}}

\def\Dbar{\leavevmode\lower.6ex\hbox to 0pt
{\hskip-.23ex\accent"16\hss}D}
\begin{document}


\def\be{\begin{eqnarray}}
\def\ee{\end{eqnarray}}


\newcommand{\ca}{\mathcal A}

\newcommand{\cb}{\mathcal B}
\newcommand{\cc}{\mathcal C}
\newcommand{\cd}{\mathcal D}
\newcommand{\ce}{\mathcal E}
\newcommand{\cf}{\mathcal F}
\newcommand{\cg}{\mathcal G}
\newcommand{\ch}{\mathcal H}
\newcommand{\ci}{\mathcal I}
\newcommand{\cj}{\mathcal J}
\newcommand{\ck}{\mathcal K}
\newcommand{\cl}{\mathcal L}
\newcommand{\cm}{\mathcal M}
\newcommand{\cn}{\mathcal N}
\newcommand{\co}{\mathcal O}
\newcommand{\cp}{\mathcal P}
\newcommand{\cq}{\mathcal Q}
\newcommand{\calr}{\mathcal R}
\newcommand{\cs}{\mathcal S}
\newcommand{\ct}{\mathcal T}
\newcommand{\cu}{\mathcal U}
\newcommand{\cv}{\mathcal V}
\newcommand{\cw}{\mathcal W}
\newcommand{\cx}{\mathcal X}
\newcommand{\cy}{\mathcal Y}
\newcommand{\cz}{\mathcal Z}


\newcommand{\sa}{\mathscr{A}}
\newcommand{\sm}{\mathscr{M}}


\newcommand{\fa}{\mathfrak{a}}  \newcommand{\Fa}{\mathfrak{A}}
\newcommand{\fb}{\mathfrak{b}}  \newcommand{\Fb}{\mathfrak{B}}
\newcommand{\fc}{\mathfrak{c}}  \newcommand{\Fc}{\mathfrak{C}}
\newcommand{\fd}{\mathfrak{d}}  \newcommand{\Fd}{\mathfrak{D}}
\newcommand{\fe}{\mathfrak{e}}  \newcommand{\Fe}{\mathfrak{E}}
\newcommand{\ff}{\mathfrak{f}}  \newcommand{\Ff}{\mathfrak{F}}
\newcommand{\fg}{\mathfrak{g}}  \newcommand{\Fg}{\mathfrak{G}}
\newcommand{\fh}{\mathfrak{h}}  \newcommand{\Fh}{\mathfrak{H}}
\newcommand{\fraki}{\mathfrak{i}}       \newcommand{\Fraki}{\mathfrak{I}}
\newcommand{\fj}{\mathfrak{j}}  \newcommand{\Fj}{\mathfrak{J}}
\newcommand{\fk}{\mathfrak{k}}  \newcommand{\Fk}{\mathfrak{K}}
\newcommand{\fl}{\mathfrak{l}}  \newcommand{\Fl}{\mathfrak{L}}
\newcommand{\fm}{\mathfrak{m}}  \newcommand{\Fm}{\mathfrak{M}}
\newcommand{\fn}{\mathfrak{n}}  \newcommand{\Fn}{\mathfrak{N}}
\newcommand{\fo}{\mathfrak{o}}  \newcommand{\Fo}{\mathfrak{O}}
\newcommand{\fp}{\mathfrak{p}}  \newcommand{\Fp}{\mathfrak{P}}
\newcommand{\fq}{\mathfrak{q}}  \newcommand{\Fq}{\mathfrak{Q}}
\newcommand{\fr}{\mathfrak{r}}  \newcommand{\Fr}{\mathfrak{R}}
\newcommand{\fs}{\mathfrak{s}}  \newcommand{\Fs}{\mathfrak{S}}
\newcommand{\ft}{\mathfrak{t}}  \newcommand{\Ft}{\mathfrak{T}}
\newcommand{\fu}{\mathfrak{u}}  \newcommand{\Fu}{\mathfrak{U}}
\newcommand{\fv}{\mathfrak{v}}  \newcommand{\Fv}{\mathfrak{V}}
\newcommand{\fw}{\mathfrak{w}}  \newcommand{\Fw}{\mathfrak{W}}
\newcommand{\fx}{\mathfrak{x}}  \newcommand{\Fx}{\mathfrak{X}}
\newcommand{\fy}{\mathfrak{y}}  \newcommand{\Fy}{\mathfrak{Y}}
\newcommand{\fz}{\mathfrak{z}}  \newcommand{\Fz}{\mathfrak{Z}}

\newcommand{\cfg}{\dot \fg}
\newcommand{\cFg}{\dot \Fg}
\newcommand{\ccg}{\dot \cg}
\newcommand{\circj}{\dot {\mathbf J}}
\newcommand{\circs}{\circledS}
\newcommand{\jmot}{\mathbf J^{-1}}


\newcommand{\rmd}{\mathrm d}
\newcommand{\mca}{\ ^-\!\!\ca}
\newcommand{\pca}{\ ^+\!\!\ca}
\newcommand{\peq}{^\Psi\!\!\!\!\!=}
\newcommand{\lt}{\left}
\newcommand{\rt}{\right}
\newcommand{\HN}{\hat{H}(N)}
\newcommand{\HM}{\hat{H}(M)}
\newcommand{\Hv}{\hat{H}_v}
\newcommand{\cyl}{\mathbf{Cyl}}
\newcommand{\lag}{\left\langle}
\newcommand{\rag}{\right\rangle}
\newcommand{\Ad}{\mathrm{Ad}}
\newcommand{\trace}{\mathrm{tr}}
\newcommand{\bbc}{\mathbb{C}}
\newcommand{\AC}{\overline{\mathcal{A}}^{\mathbb{C}}}
\newcommand{\Ar}{\mathbf{Ar}}
\newcommand{\uc}{\mathrm{U(1)}^3}
\newcommand{\M}{\hat{\mathbf{M}}}
\newcommand{\spin}{\text{Spin(4)}}
\newcommand{\id}{\mathrm{id}}
\newcommand{\Pol}{\mathrm{Pol}}
\newcommand{\Fun}{\mathrm{Fun}}
\newcommand{\bp}{p}
\newcommand{\act}{\rhd}
\newcommand{\data}{\lt(j_{ab},A,\bar{A},\xi_{ab},z_{ab}\rt)}
\newcommand{\datao}{\lt(j^{(0)}_{ab},A^{(0)},\bar{A}^{(0)},\xi_{ab}^{(0)},z_{ab}^{(0)}\rt)}
\newcommand{\deltadata}{\lt(j'_{ab}, A',\bar{A}',\xi_{ab}',z_{ab}'\rt)}
\newcommand{\background}{\lt(j_{ab}^{(0)},g_a^{(0)},\xi_{ab}^{(0)},z_{ab}^{(0)}\rt)}
\newcommand{\sgn}{\mathrm{sgn}}
\newcommand{\vth}{\vartheta}
\newcommand{\rmi}{\mathrm{i}}
\newcommand{\bfmu}{\pmb{\mu}}
\newcommand{\bfnu}{\pmb{\nu}}
\newcommand{\bfm}{\mathbf{m}}
\newcommand{\bfn}{\mathbf{n}}


\newcommand{\sz}{\mathscr{Z}}
\newcommand{\sk}{\mathscr{K}}

\title{Perturbational Decomposition Analysis for Quantum Ising Model with Weak Transverse Fields}

\author{Youning Li}%
\affiliation{College of Science, China Agriculture University, Beijing, People's Republic of China}
\author{Junfeng Huang}
\affiliation{College of Science, China Agriculture University, Beijing, People's Republic of China}
\author{Chao Zhang}
\affiliation{Department of Physics, The Hong Kong University of Science and Technology, Clear Water Bay, Kowloon, Hong Kong}
\author{Jun Li}
\affiliation{Institute of Quantum Precision Measurement, State Key Laboratory of Radio
Frequency Heterogeneous Integration, Shenzhen University, Shenzhen 518060, China}%
\affiliation{College of Physics and Optoelectronic Engineering, Shenzhen University, Shenzhen 518060, China}

%

\date{\today}

\begin{abstract}
This work presented a perturbational decomposition method for simulating quantum evolution under the one-dimensional Ising model with both longitudinal and transverse fields. By treating the transverse field terms as perturbations in the expansion, our approach is particularly effective in systems with moderate longitudinal fields and weak to moderate transverse fields relative to the coupling strength. Through systematic numerical exploration, we characterized parameter regimes and evolution time windows where the decomposition achieved measurable improvements over conventional Trotter decomposition methods. The developed perturbational approach and its characterized parameter space may provide practical guidance for choosing appropriate simulation strategies in different parameter regimes of the one-dimensional Ising model.
\end{abstract}

\maketitle
\renewcommand\theequation{\arabic{section}.\arabic{equation}}
\setcounter{tocdepth}{4}
\makeatletter
\@addtoreset{equation}{section}
\makeatother

\section{Introduction}
Quantum simulation has emerged as a crucial tool for understanding complex quantum systems, with applications ranging from quantum chemistry to condensed matter physics\cite{childs2019nearly, cao2019quantum, altman2021quantum}. Recent work has demonstrated significant theoretical advances in simulation algorithms \cite{childs2021theory, haah2021quantum, tran2021faster}, while experimental achievements have shown the potential of quantum simulators in various platforms\cite{semeghini2021probing, ebadi2021quantum,mi2022time}, highlighting their practical significance in exploring quantum phenomena. In most physical systems of interest, the Hamiltonian can be expressed as a sum of local terms:

\begin{equation}
H = \sum_{k=1}^{M} H_k,
\end{equation}

\noindent where each $H_k$ acts non-trivially only on a limited number of degrees of freedom. This local structure naturally leads to simulation strategies based on the Trotter-Suzuki decomposition. The theoretical foundation was established by Suzuki's pioneering work on generalized Trotter formulas\cite{Suzuki1976generalized}, further developed through quantum analysis methods\cite{Suzuki1997quantum}, and extended to higher-order product formulas\cite{Hatano2005Finding}. The most basic form of such decomposition approximates\cite{lloyd1996universal,trotter1959product} the evolution operator as:

\begin{equation}
e^{-itH} \approx \left(e^{-it H_1/n} e^{-it H_2/n} \cdots e^{-it H_M/n}\right)^n,
\end{equation}

\noindent where $n$ is the number of Trotter steps.

Conventional quantum simulation methods typically maintain a crucial constraint: the sum of coefficients in the decomposed terms exactly matches the original Hamiltonian generating the evolution. This approach, while mathematically natural, may not always yield optimal simulation accuracy. Recent theoretical developments have further expanded our understanding of efficient product formulas and their applications\cite{childs2019nearly, tran2021faster, childs2021theory}. Higher-order decomposition schemes have been developed to improve simulation accuracy\cite{Hatano2005Finding}, though their effectiveness can be limited by the increasing complexity of implementation. Recent work has shown that careful consideration of the system's structure and parameter regimes can lead to more efficient simulation strategies\cite{low2019hamiltonian, roetteler2017quantum,low2018hamiltonian}.

A particularly important model in quantum many-body physics is the quantum Ising model with both longitudinal and transverse fields\cite{coldea2010quantum, sachdev1999quantum,dutta2015quantum}. This model not only serves as a paradigmatic example for studying quantum phase transitions but also represents a practical testing ground for quantum simulation techniques\cite{bernien2017probing,zhang2017observation,king2018observation}. The quantum Ising model has been extensively studied both theoretically and experimentally\cite{simon2011quantum,lanyon2011universal,struck2011quantum}, providing crucial insights into quantum criticality and many-body phenomena.

The applicability of Ising-type models extends far beyond traditional condensed matter physics. In biological physics, these models have been successfully applied to understand protein folding dynamics and to infer protein structure from sequence data\cite{ferguson2013translating,morcos2011direct,weigt2009identification}. In collective behavior studies, they have helped explain the emergence of coordinated motion in biological systems\cite{cavagna2010scale}. The model has also found applications in studying complex biological networks and their dynamics\cite{bialek2012statistical,mora2016local}.

Recent developments in machine learning and artificial intelligence have further expanded the reach of Ising-type models\cite{carleo2017solving, mehta2019high,torlai2018neural}. These connections have not only provided new computational tools for studying quantum many-body systems but have also established bridges between quantum physics and modern data science. In many of these applications, the transverse field terms appear as perturbations to the primary Ising interactions, creating a natural hierarchy in the energy scales of the system.

Building upon the perturbation composition framework, this work presented a decomposition approach that deliberately deviates from the conventional coefficient-matching constraint. We focused specifically on simulating quantum evolution under the one-dimensional Ising model, treating transverse field strength as perturbations in the expansion. Our approach exploits the natural parameter hierarchy in systems where the transverse field strength is moderate relative to the coupling strength. For certain evolution time ranges and within appropriate parameter regimes, our method achieves measurable improvements in simulation fidelity compared to conventional approaches.

Our numerical exploration mapped out the evolution of time windows across different field strength combinations, providing a comprehensive visualization of where this method offers noticeable improvements over conventional decomposition techniques. This systematic investigation not only extends recent theoretical understanding of Trotter decomposition\cite{childs2021theory} to practical quantum simulations, but also reveals the practical boundaries and limitations of our perturbational approach within the parameter space of the one-dimensional Ising model.

The remainder of this paper is organized as follows. In Section II, we review the fundamental concepts and establish our notation. Section III presents our main analysis of the perturbational decomposition method and its optimization results. Finally, we conclude with a discussion of the implications and potential applications of our results.

\section{Preliminary}\label{II}
We begin by examining the optimization of unitary evolution simulation under a single-qubit Hamiltonian:
\begin{equation}\label{qubit}
  H=\alpha \sigma_x+ \sigma_z,
\end{equation}
where $\alpha$ represents the relative strength of the transverse field.

The Suzuki-Trotter decomposition\cite{Suzuki1976generalized} provides a fundamental approach to simulating quantum evolution. Its simplest form, known as the second-order Trotter formula, gives:
\begin{equation}\label{trotter}
  e^{-it\alpha \sigma_x-it \sigma_z}\approx e^{-it\frac{\alpha}{2} \sigma_x} e^{-it\sigma_z} e^{-it\frac{\alpha}{2} \sigma_x}+O(t^3).
\end{equation}

The accuracy of this decomposition deteriorates for finite time intervals due to the non-commuting nature of $\sigma_x$ and $\sigma_z$. This motivates us to introduce an optimization parameter $\lambda$ to improve the simulation fidelity.

Let $U=e^{-it\alpha \lambda \sigma_x-it \sigma_z}$ denote the target evolution and $V=e^{-it\frac{\alpha}{2} \sigma_x}e^{-it\sigma_z} e^{-it\frac{\alpha}{2} \sigma_x}$ represent the Trotter decomposition. The optimization problem can be formulated as:
\begin{equation}\label{fidelity}
  \max\limits_{\lambda}|\textrm{Tr}(UV^{\dagger})|^2/2^2.
\end{equation}
This optimization can be achieved either by direct expansion or by differentiating both sides with respect to $\alpha$ and equating them:
\begin{subequations}\label{dphi}
\begin{align}
  \frac{d}{d\alpha}(e^{-it\alpha \lambda \sigma_x-it \sigma_z})&=\frac{d e^{\Phi}}{d \Phi}\frac{\partial \Phi}{\partial \alpha},   \\
  \frac{d}{d\alpha}(e^{-it\frac{\alpha}{2} \sigma_x} e^{-it \sigma_z} e^{-it\frac{\alpha}{2} \sigma_x}) &=
  \frac{-it}{2}(\sigma_x e^{\Phi}+e^{\Phi}\sigma_x)=\frac{-it}{2}e^{\Phi}(e^{-\Phi}\sigma_xe^{\Phi}+1)\sigma_x,
\end{align}
\end{subequations}
where $\Phi(\alpha)=-it\alpha \lambda \sigma_x-it\sigma_z$.

To proceed, we need to calculate $\frac{df(A(\alpha))}{d \alpha}$, taking into account the non-zero commutator between $\Phi(\alpha)$ and $d\Phi(\alpha)$. Following \cite{Suzuki1997quantum,Hatano2005Finding}, we briefly review the key steps:

${d f(A(\alpha))}/{d \alpha}$ must be expressed in terms of $A$ and the commutation relation of $A$, or the "inner derivation":
\begin{equation}\label{inner deriv}
\delta_A \equiv [A, ].
\end{equation}

It can be directly obtained from the above definition:
\begin{subequations}\label{inner}
\begin{align}
  \delta_{aA+bB} & =a \delta_A +b \delta_B, \\
  [A^m,\delta_{A^n}] &=0.
\end{align}
\end{subequations}

By induction of the power, one may obtain
\begin{subequations}\label{delta_An}
\begin{align}
  \delta_{A^n} &=A^n-(A-\delta_A)^n,  \\
  e^{xA}Be^{-xA}&=e^{x\delta_A}B,\\
  dA^n &=\lim_{h\rightarrow 0} \frac{(A+h dA)^n-A^n}{h}=(A^n-(A-\delta_A)^n)\delta_A dA.
\end{align}
\end{subequations}
Therefore,
\begin{equation}\label{dAn}
  \frac{d(A^n)}{dA}=\frac{A^n-(A-\delta_A)^n}{\delta_A}=\frac{\delta_{A^n}}{\delta_A},
\end{equation}
Here, while $\delta_A$ in the denominator appears formally ill-defined, it cancels out during the expansion of the numerator.
By Taylor expansion, Eq(\ref{dAn}) could be generalized to
\begin{equation}\label{df}
  \frac{df(A)}{dA}=\frac{f(A)-f(A-\delta_A)}{\delta_A}=\frac{\delta_{f(A)}}{\delta_A}.
\end{equation}
For the special case $f(A)=e^A$, Eq(\ref{df}) reads
\begin{equation}\label{deA}
  \frac{de^A}{dA}=\frac{e^A-e^{A-\delta_A}}{\delta_A}=e^A\frac{1-e^{-\delta_A}}{\delta_A}.
\end{equation}

Substituting Eq(\ref{dAn}) and Eq(\ref{deA}) into Eq(\ref{dphi}) and equating them obtain:
\begin{equation}\label{alpha}
  \frac{\partial \Phi}{\partial \alpha}=\frac{-it}{2}\delta_{\Phi}\frac{1+e^{-\delta_{\Phi}}}{1-e^{-\delta_{\Phi}}}\sigma_x.
\end{equation}

In the perturbative limit $\alpha\ll 1$, we obtain
\begin{equation}\label{lambdaequation}
  \lambda \sigma_x=\frac{-it\sigma_z}{2}\frac{1+\exp(it\delta_{\sigma_z})}{1-\exp{it\delta_{\sigma_z}}}\sigma_x,
\end{equation}
The optimization yields the optimal parameter:
\begin{equation}\label{lambdavalue}
  \lambda=\frac{t}{\tan(t)},
\end{equation}
which provides enhanced simulation accuracy for finite evolution times.

After relabeling the parameters, we can express the unitary evolution for single-qubit rotation in a more compact form. Our perturbative method yields:
\begin{equation}\label{perturbative_single}
  e^{-it\alpha' \sigma_x-it \sigma_z}\approx e^{-it\frac{\alpha'}{2}\frac{\tan t}{t} \sigma_x} e^{-it\sigma_z} e^{-it\frac{\alpha'}{2} \frac{\tan t}{t} \sigma_x}+O(\alpha'^2).
\end{equation}

We compare the performance of our perturbative method with the traditional Trotter decomposition shown in Eq.(\ref{trotter}). As demonstrated in Fig. 1, the perturbative method matches the performance of Trotter decomposition in the small $t$ regime while exhibiting superior fidelity for larger evolution times $t$.

\begin{figure}[h!]
\centering
\includegraphics[width=0.5\textwidth]{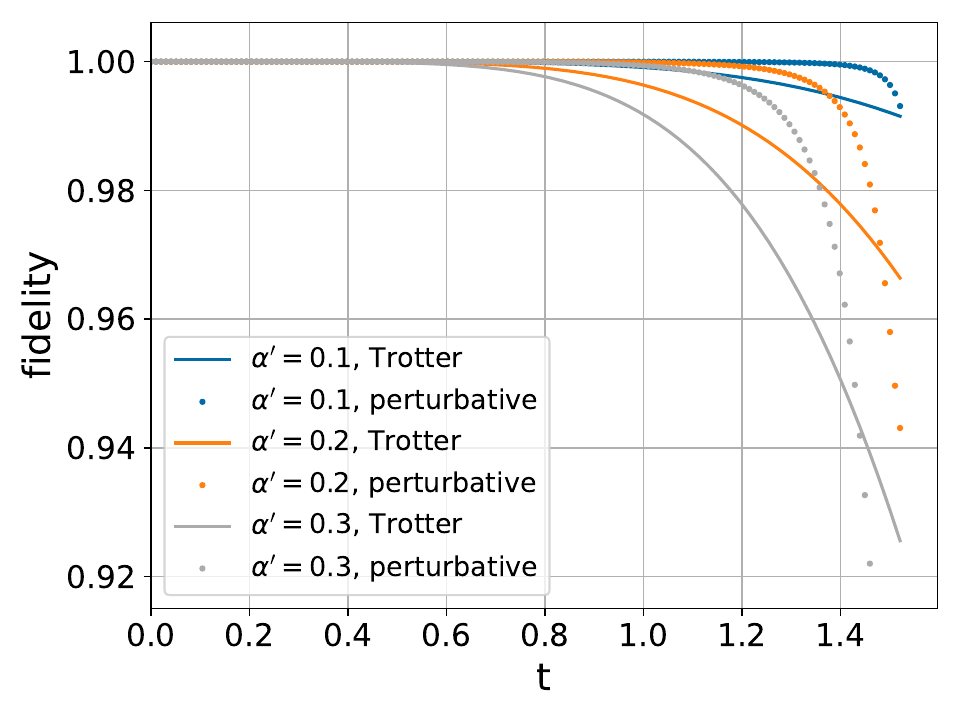}
\caption{Comparison of fidelity between the traditional Trotter decomposition and our perturbative expansion method.}
\end{figure}

\section{Perturbational Composition for Ising Model}

Extending our previous analysis of single-qubit optimization, we now investigate a more complex system: a one-dimensional Ising chain with periodic boundary conditions under both longitudinal and transverse fields. The Hamiltonian of this system can be expressed as:

\begin{equation}
H=\sum_{k=1}^{2n} J_{k,k+1}Z^{(k)}Z^{(k+1)}+g_k Z^{(k)}+h_k X^{(k)},
\end{equation}

where $J_{k,k+1}$ represents the nearest-neighbor coupling strength, while $g_k$ and $h_k$ denote the longitudinal and transverse fields respectively. We focus on the regime where $h_k \ll J_{k,k+1}, g_k$. The system satisfies periodic boundary conditions through $2n+1 \equiv 1$.

This Hamiltonian naturally decomposes into four distinct components:
\begin{subequations}
\begin{eqnarray}
A_1 & = & \sum_{k=1}^{n}J_{2k-1,2k}Z^{(2k-1)}Z^{(2k)},\\
A_2 & = & \sum_{k=1}^{n}J_{2k,2k+1}Z^{(2k)}Z^{(2k+1)},\\
B_1 & = & \sum_{k=1}^{2n}g_k Z^{(k)},\\
B_2 & = & \sum_{k=1}^{2n}h_k X^{(k)}.
\end{eqnarray}
\end{subequations}

Thus, we have $H=A_1+A_2+B_1+B_2$. However, unlike the simpler case discussed in Section \ref{II}, we cannot simultaneously remove $A_1+A_2+B_1$. This limitation arises because $\delta_{A_1+A_2+B_1}^2B_2$ inevitably generates terms such as $Z^{(k-1)}X^{(k)}Z^{(k+1)}$ or $Z^{(k-1)}X^{(k)}+X^{(k-1)}Z^{(k)}$, necessitating a sequential removal approach.

\vspace{1em}

To address this new challenge in many-body systems, let us first examine a simplified but illustrative example. Consider the exponential product form:
\[
e^{-it\left(J_{1,2}Z^{(1)}Z^{(2)}+g_1Z^{(1)}+g_2Z^{(2)}+h_1X^{(1)}+h_2X^{(2)}\right)}.
\]

Following the methodology of \cite{Hatano2005Finding}, we analyze:
\begin{eqnarray}\label{Phi}
e^{-i\frac{t}{2}\alpha(B+C)}e^{-it A} e^{-\frac{t}{2}\alpha(B+C)}=e^{-i\Phi(\alpha, t)}=e^{-it(A+\alpha p(t)+O(\alpha^2))},
\end{eqnarray}
where we define $A=J_{1,2}Z^{(1)}Z^{(2)}$, $B=g_1Z^{(1)}+g_2Z^{(2)}$, and $C=h_1X^{(1)}+h_2X^{(2)}$, with $\alpha$ serving as a small parameter.

By differentiating both sides of Eq.(\ref{Phi}) with respect to $\alpha$, we obtain:
\begin{equation}
\frac{d}{d\alpha}e^{-it(A+\alpha p(t)+O(\alpha^2))}
=\frac{de^{\Phi}}{d\Phi}\frac{\partial \Phi(\alpha, t)}{\partial \alpha}
=e^{\Phi}\frac{1-e^{-\delta_{\Phi}}}{\delta_{\phi}}\frac{\partial \Phi}{\partial \alpha},
\end{equation}

\begin{eqnarray}
\frac{d}{d\alpha}e^{-i\frac{t}{2}\alpha(B+C)}e^{-it A} e^{-i\frac{t}{2}\alpha(B+C)}
&=&\frac{-it}{2}e^{\Phi}\left(e^{-\Phi}(B+C)e^{\Phi}+(B+C)\right)\notag\\
&=&\frac{-it}{2}e^{\Phi}(e^{-\delta_{\Phi}}+1)(B+C).
\end{eqnarray}

Equating these expressions yields:
\begin{eqnarray}
\frac{\partial \Phi}{\partial \alpha}=\frac{-it}{2}\delta_{\Phi}\frac{1+e^{-\delta_{\Phi}}}{1-e^{-\delta_{\Phi}}}(B+C),
\end{eqnarray}

Direct calculation of $\frac{\partial\Phi}{\partial \alpha}$ gives:
\begin{equation}
\frac{\partial\Phi}{\partial \alpha}=-it p(t).
\end{equation}

Consequently:
\begin{equation}
p(t)=\frac{1}{2}\delta_{\Phi}\frac{1+e^{-\delta_{\Phi}}}{1-e^{-\delta_{\Phi}}}(B+C),
\end{equation}

For $\alpha \ll 1$, we can obtain a Taylor expansion:
\begin{equation}
p(t)=\frac{-it\delta_A}{2}\frac{1+e^{it\delta_A}}{1-e^{it\delta_A}}(B+C)=\frac{1}{2}\sum_{k=0}^{\infty}a_k (-it)^k \delta_{A}^k (B+C),
\end{equation}
where $a_0=2$.

Through careful analysis, we find:
\begin{subequations}
\begin{eqnarray}
\delta_A(B+C)&=&[A,B+C]=[A,C]=2iJ_{1,2}(h_1Y^1Z^2+h_2Z^1Y^2),\\
\delta_A^2(B+C)&=&2iJ_{1,2}[A,h_1Y^1Z^2+h_2Z^1Y^2]=4J_{1,2}^2(h_1X^1+h_2X^2)=4J_{1,2}^2C,\\
\delta_A^3(B+C)&=&4[A,C]=8iJ_{1,2}^3(h_1Y^1Z^2+h_2Z^1Y^2),\\
&&\cdots.
\end{eqnarray}
\end{subequations}

Moreover, since $p(t)$ satisfies the symmetry relation $p(-t)=p(t)$, we can neglect all odd terms in the expansion, yielding:
\begin{equation}
p(t)=\frac{1}{2}\left(2B+\sum_{k=0}^{\infty}a_k(-2it)^k\right)=B+\frac{1}{f(J_{1,2}t)}C,
\end{equation}
where $f(t)=\frac{\tan(t)}{t}$ is the scaling function.

This leads to the following equivalent expression:
\begin{eqnarray}\label{perturbative_2qubit}
e^{-it(A+\alpha' B + \alpha' C)}=e^{-i\frac{t}{2}\alpha'(B+f(J_{1,2}t)C)}e^{-it A} e^{-\frac{t}{2}\alpha'(B+f(J_{1,2}t)C)}.
\end{eqnarray}
\vspace{1em}

Following the same procedure iteratively, we obtain the optimized exponential product formulas:
\begin{subequations}
\begin{eqnarray}
e^{-it\left(A_1+ A_2+ B_1+  B_2 \right)}
&\approx& e^{-i\frac{t}{2}(A_2+B_1+B^{(1)}_2)}e^{-itA_1}e^{-i\frac{t}{2}(A_2+B_1+B^{(1)}_2)} ,\\
&\approx& e^{-i\frac{t}{4}(B_1+B^{(2)}_2)}e^{-i\frac{t}{2}A_2}e^{-i\frac{t}{4}(B_1+B^{(2)}_2)}e^{-itA_1}
e^{-i\frac{t}{4}(B_1+B^{(2)}_2)}e^{-i\frac{t}{2}A_2}e^{-i\frac{t}{4}(B_1+B^{(2)}_2)}\\
&\approx&
e^{-i\frac{t}{8}B^{(3)}_2}e^{-i\frac{t}{4} B_1}e^{-i\frac{t}{8}B^{(3)}_2}e^{-i\frac{t}{2}A_2}\notag\\
&&\times
e^{-i\frac{t}{8}B^{(3)}_2}e^{-i\frac{t}{4} B_1}e^{-i\frac{t}{8}B^{(3)}_2}e^{-itA_1} e^{-i\frac{t}{8}B^{(3)}_2}e^{-i\frac{t}{4} B_1}e^{-i\frac{t}{8}B^{(3)}_2}\notag\\
&&\times e^{-i\frac{t}{2}A_2}e^{-i\frac{t}{8}B^{(3)}_2}e^{-i\frac{t}{4} B_1}e^{-i\frac{t}{8}B^{(3)}_2}\label{perturbative_nqubit},
\end{eqnarray}
\end{subequations}
where the modified operators $B_2^{(k)}$ incorporate the optimization parameters:
\begin{subequations}
\begin{eqnarray}
B_2^{(1)}&=&\sum_{k=1}^{2n}f(J_{2k-1,2k}t)h_{2k-1}X^{(2k-1)}+f(J_{2k-1,2k}t)h_{2k}X^{(2k)},\\
B_2^{(2)}&=&\sum_{k=1}^{2n}f(J_{2k,2k+1}t)f(J_{2k-1,2k}t)h_{2k}X^{(2k)}+f(J_{2k,2k+1}t)f(J_{2k+1,2k+2}t)h_{2k+1}X^{(2k+1)},\\
B_2^{(3)}&=&\sum_{k=1}^{2n}f(g_{2k}t)f(J_{2k,2k+1}t)f(J_{2k-1,2k}t)h_{2k}X^{(2k)}+f(g_{2k+1})f(J_{2k,2k+1}t)f(J_{2k+1,2k+2}t)h_{2k+1}X^{(2k+1)}\notag\\
         &=&\sum_{k=1}^{2n} f(g_k t)f(J_{k-1,k}t)f(J_{k,k+1}t)h_k X^{(k)}.\label{B23}
\end{eqnarray}
\end{subequations}

Notably, Eq(\ref{perturbative_nqubit}) provides a more accurate expansion compared to the conventional second-order Trotter expansion:
\begin{equation}\label{2-trotter}
  e^{-it\left(A_1+ A_2+ B_1+  B_2 \right)} \approx e^{-i\frac{t}{2} B_2} e^{-it\left(A_1+ A_2+ B_1 \right)} e^{-i\frac{t}{2} B_2}.
\end{equation}

\begin{figure}[h!]
\centering
\includegraphics[width=0.5\textwidth]{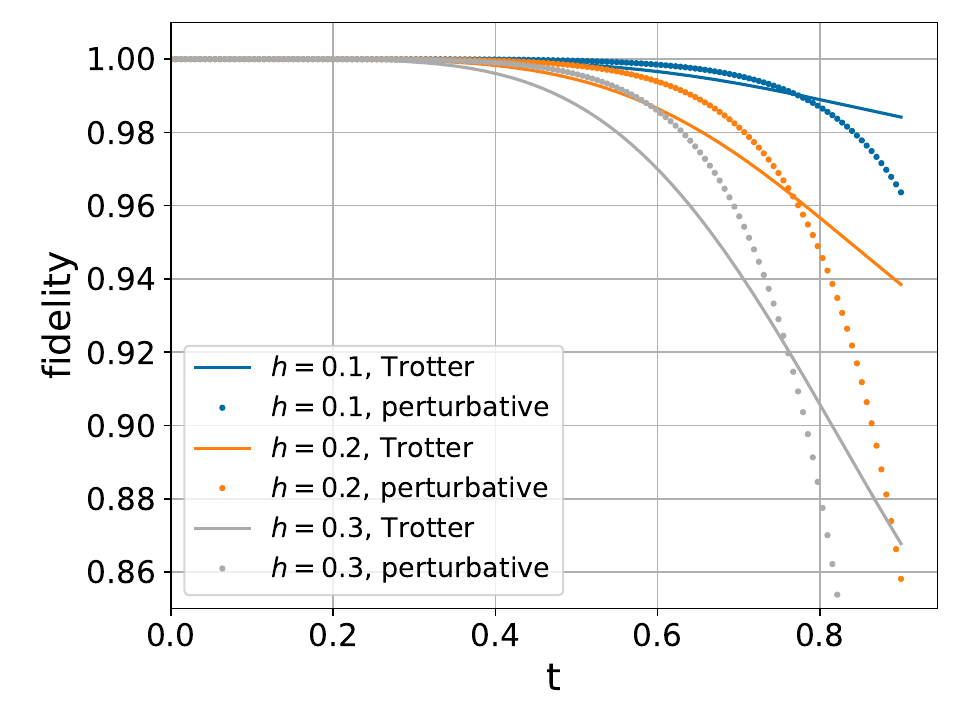}
\caption{Fidelity comparison between second-order Trotter decompositionin in Eq(\ref{2-trotter}) and perturbative expansion in Eq(\ref{perturbative_nqubit}) for a $6$-spin chain. The coupling strengths are set to $J_{k,k+1}=1$ (uniform nearest-neighbor interaction), with longitudinal field $g_k=1$ and transverse field $h_k=h$ applied to each spin.}
\end{figure}

While our perturbative method achieves higher accuracy, it requires slightly more than three times ($3.375\times$) as many local unitary operations. Therefore, a fair comparison necessitates careful evaluation of the trade-off between computational cost and simulation accuracy.

By setting the composite coefficient $f(g_k t)f(J_{k-1,k}t)f(J_{k,k+1}t)$ to unity in each term of Eq(\ref{B23}), we recover a higher-order Trotter expression. Our approach can thus be viewed as a variational optimization of the higher-order Trotter coefficients, providing a natural framework for performance comparison between optimized (variational) and fixed (unit) coefficients.

\begin{figure}[h!]
\centering
\includegraphics[width=0.5\textwidth]{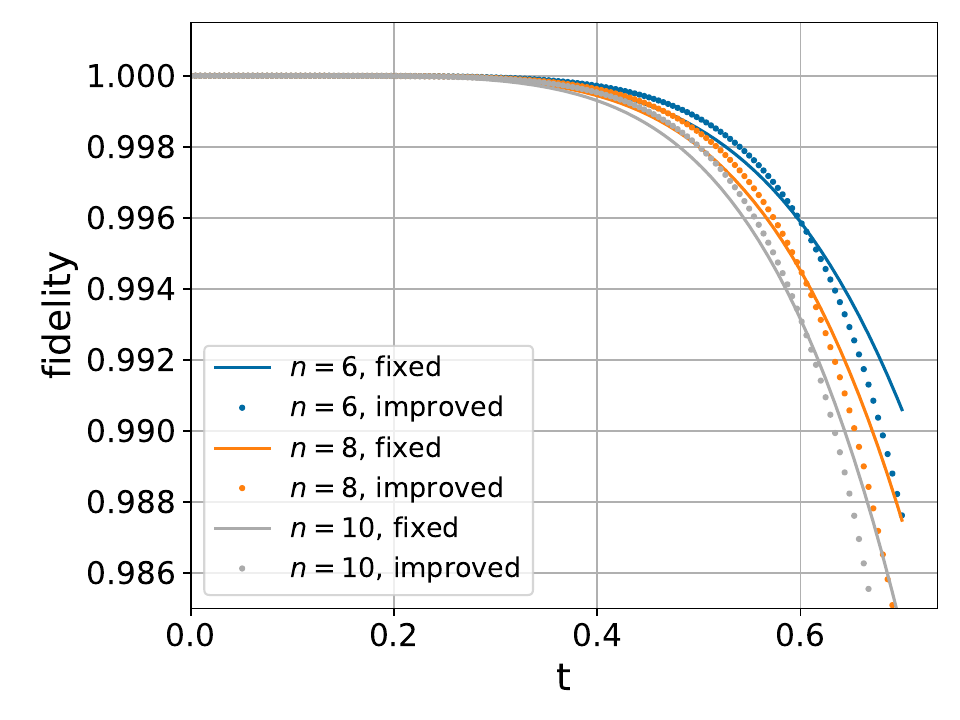}
\caption{Fidelity comparison between fixed and optimized coefficients for varying chain lengths ($2n=6,8,10$). Parameters: uniform coupling $J_{k,k+1}=1$, longitudinal field $g_k=0.2$, and transverse field $h_k=0.3$.}
\label{fig:size_comparison}
\end{figure}

We first investigate the dependence on system size. As shown in Fig.~\ref{fig:size_comparison}, the performance characteristics remain qualitatively similar across different chain lengths, justifying our choice of a $6$-spin system for subsequent numerical analysis.

\begin{figure}[h!]
\centering
\includegraphics[width=0.5\textwidth]{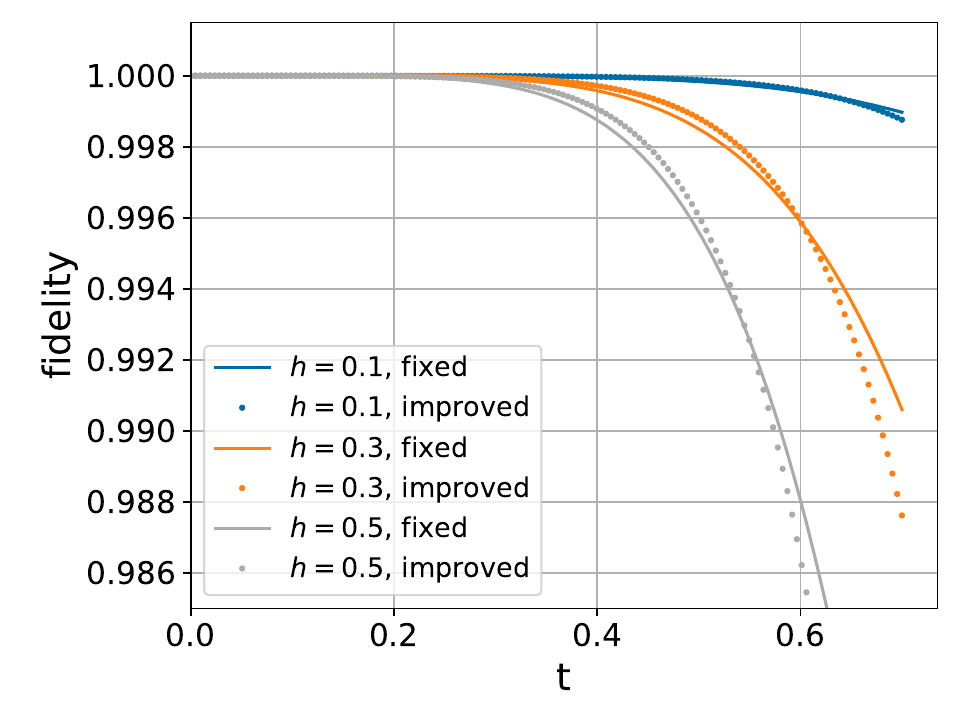}
\caption{Performance comparison for varying transverse field strengths ($h$) in a $6$-spin chain. Fixed parameters: $J_{k,k+1}=1$, $g_k=0.2$.}
\label{fig:field_comparison}
\end{figure}

In realistic physical systems, the ratio $g_k/J_{l,l+1}$ typically assumes values around $0.2$. Within this regime, Fig.~\ref{fig:field_comparison} demonstrates that our perturbative approach achieves noticeably improved fidelity compared to the higher-order Trotter expansion for moderate transverse field strengths.

\begin{figure}[h!]
\centering
\includegraphics[width=0.5\textwidth]{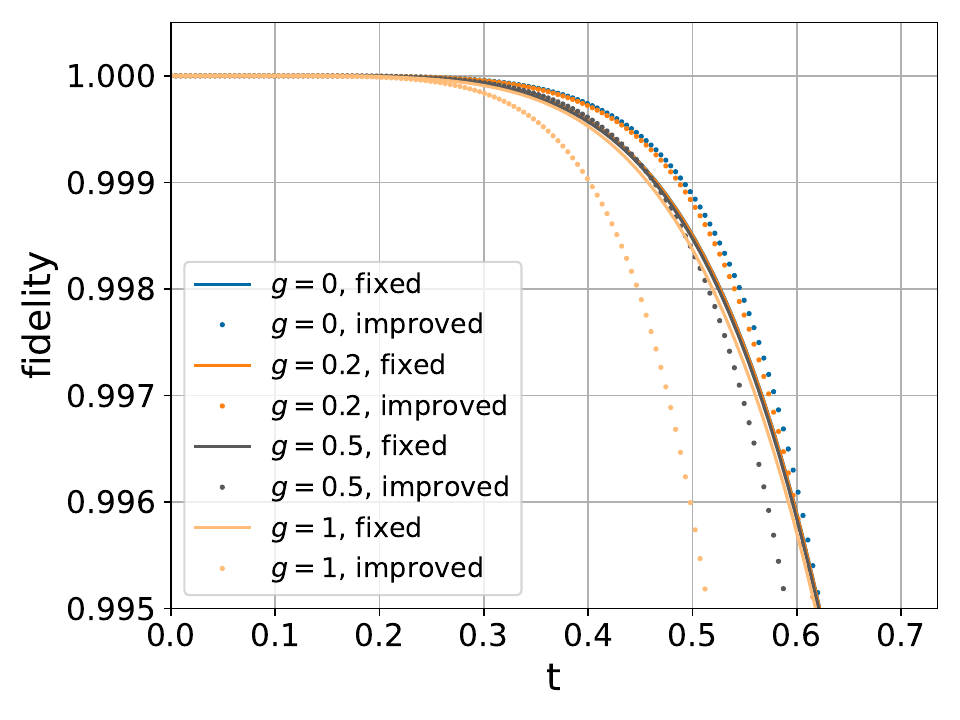}
\caption{Effect of longitudinal field strength ($g$) on simulation fidelity in a $6$-spin chain. Fixed parameters: $J_{k,k+1}=1$, $h_k=0.3$.}
\label{fig:long_field}
\end{figure}

However, this advantage diminishes and eventually reverses as the longitudinal field strength increases, as illustrated in Fig.~\ref{fig:long_field}.

\begin{figure}[h!]
\centering
\includegraphics[width=0.5\textwidth]{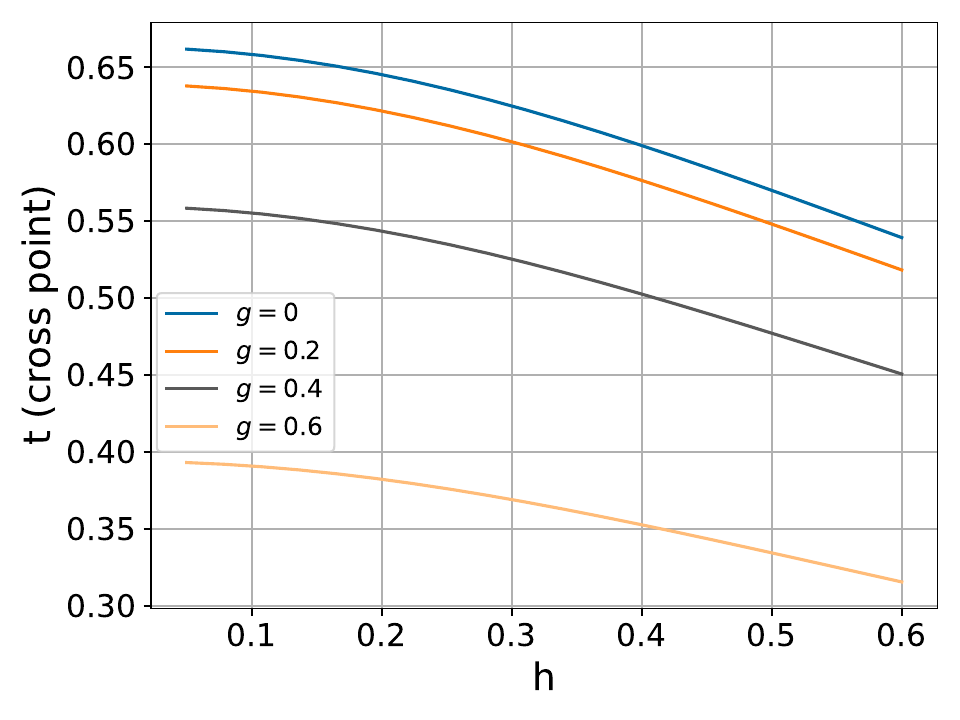}
\caption{Temporal regions where the perturbative approach achieves higher fidelity than the Trotter decomposition. Fixed parameters: nearest-neighbor coupling $J_{k,k+1}=1$, transverse field $h_k=0.3$.}
\label{fig:twindow_long}
\end{figure}

Detailed numerical analysis reveals parameter regimes where the perturbative method achieves improved fidelity compared to the Trotter decomposition. Fig.~\ref{fig:twindow_long} illustrates the time windows of this improvement across different longitudinal field strengths.

\begin{figure}[h!]
\centering
\includegraphics[width=0.5\textwidth]{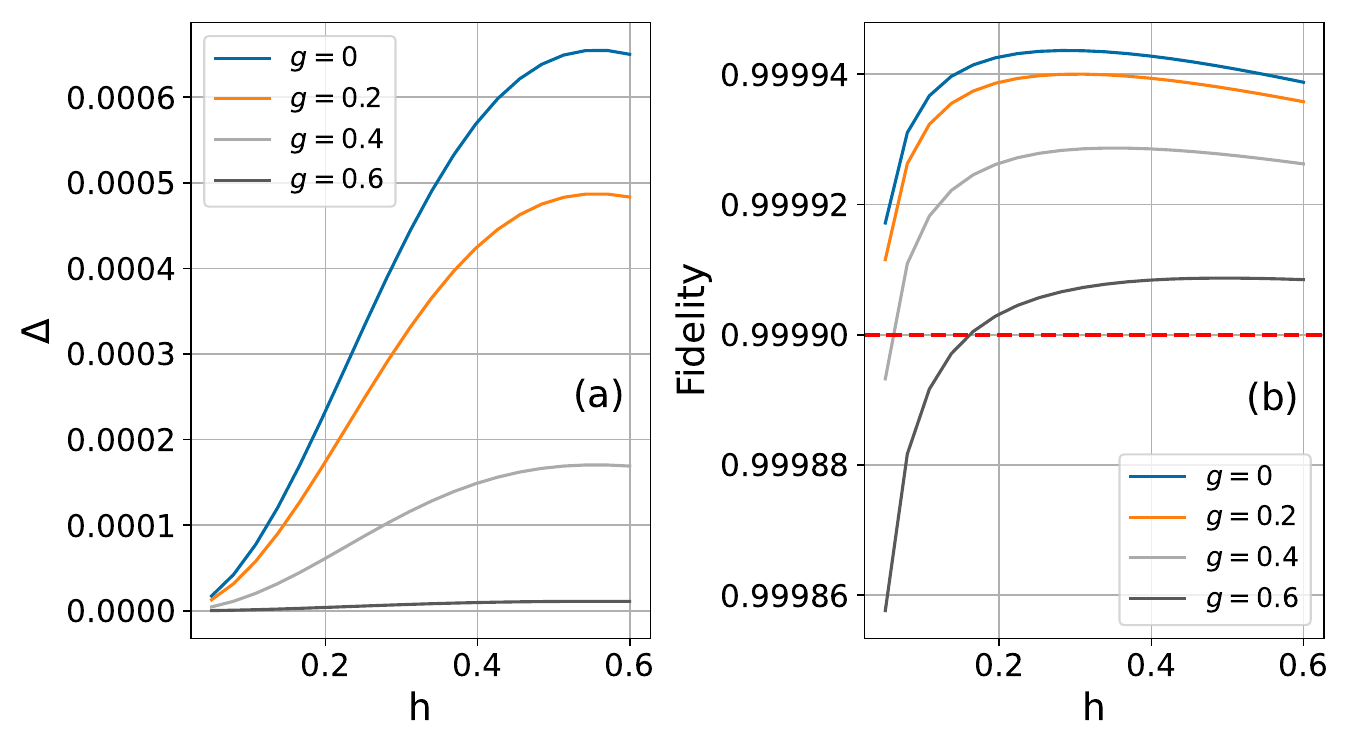}
\caption{Performance analysis of the perturbative approach in a $6$-spin chain with fixed nearest-neighbor coupling $J_{k,k+1}=1$. (a) Maximum fidelity enhancement of perturbative approach compared to the Trotter decomposition for various field strengths within the superior time window. (b) Fidelity of the perturbative approach measured at time points where the Trotter decomposition achieves 0.9999 fidelity, shown as a function of longitudinal field strength.}
\label{fig:effectiveness}
\end{figure}

The comparative performance is quantified in Fig.~\ref{fig:effectiveness}. Panel (a) shows the maximum fidelity improvement of the perturbative approach relative to the Trotter decomposition in the time window obtained in Fig.~\ref{fig:twindow_long}. Panel (b) presents the perturbative method's fidelity at time intervals where the Trotter decomposition achieves a fidelity of 0.9999. At small longitudinal field strengths, our perturbative approach reduces simulation errors by 40\% to 60\% compared to the Trotter decomposition, evaluated at a baseline error around $10^{-4}$ (0.9999 fidelity). Such improvement could be meaningful for quantum computing applications, particularly in intermediate-scale quantum devices where error accumulation is a critical concern. However, beyond a certain longitudinal field strength, the advantages of the perturbative approach become practically negligible.

\section{Conclusion}

This work presented a comprehensive analysis and extension of quantum simulation schemes based on perturbational decomposition. Our investigation revealed the following results.

We extended single-qubit optimization techniques to many-body systems by treating transverse field strength as perturbations in the expansion.
Our numerical exploration mapped out the evolution time windows across a continuous range of longitudinal and transverse field strengths, revealing systematic trends in the parameter space.

The central achievement lay in our application of perturbational decomposition to the one-dimensional Ising chain with competing longitudinal and transverse fields. Through optimization of the decomposition, we achieved noticeable improvements in simulation fidelity compared to conventional approaches for systems with typical coupling ratios ($g_k/J_{k,k+1} \approx 0.2$).

However,this advantage exhibits clear parameter boundaries. The method performs optimally in weak transverse field regimes, with time windows systematically shrinking as transverse field strength increases. While improvements persist at higher longitudinal field strengths, the relative enhancement becomes marginal beyond certain thresholds.

These findings contribute to both theoretical understanding and practical implementation of quantum simulation techniques. Our method offers a well-characterized tool for precise evolution within specific, physically relevant parameter regimes of the one-dimensional Ising model. The identification of favorable parameter regimes and evolution time windows provides useful guidance for practical applications. We expect our perturbational decomposition strategy to inspire new simulation approaches for other quantum many-body systems.

\section*{Acknowledgments}
Y-N.Li is supported by National Natural Science Foundation of China under Grant No. 12005295.
J.L is supported by National Natural Science Foundation of China (Grants No. 12122506), Shenzhen Science and Technology Program (Grant No. RCYX20200714114522109), Guangdong Basic and Applied Basic Research Foundation (Grant No. 2021B1515020070).

\end{document}